\documentclass{ieeeaccess}
\usepackage{cite}
\usepackage{amsmath,amssymb,amsfonts}
\usepackage{algorithmic}
\usepackage{graphicx}
\usepackage{textcomp}

\let\emph\textit

\def\BibTeX{{\rm B\kern-.05em{\sc i\kern-.025em b}\kern-.08em
    T\kern-.1667em\lower.7ex\hbox{E}\kern-.125emX}}
\begin{document}
\doi{N/A}

\title{Automatic Lumbar Spinal CT Image Segmentation with a Dual Densely Connected U-Net}
\author{\uppercase{He Tang}\authorrefmark{1},
\uppercase{Xiaobing Pei}\authorrefmark{1},
\uppercase{Shilong Huang}\authorrefmark{2},
\uppercase{Xin Li}\authorrefmark{3},
\uppercase{and Chao Liu}\authorrefmark{1}}
\address[1]{School of Software, Huazhong University of Science and Technology, Luoyu Road 1037, Wuhan, 430074, China}
\address[2]{Department of Orthopedics, Tongji Hospital, Huazhong University of Science and Technology, Jiefang Road 1095, Wuhan, 430030, China}
\address[3]{Department of Pediatric, Union Hospital, Huazhong University of Science and Technology, Jiefang Road 1277, Wuhan, 430030, China}
\tfootnote{This work was supported by the National Natural Science Foundation of China grant 61902139 and 81070691.}

\markboth
{He Tang \headeretal: Preparation of Papers for IEEE TRANSACTIONS and JOURNALS}
{He Tang \headeretal: Preparation of Papers for IEEE TRANSACTIONS and JOURNALS}

\corresp{Corresponding author: Shilong Huang (e-mail: doctorhsl@163.com).}

\begin{abstract}
The clinical treatment of degenerative and developmental lumbar spinal stenosis (LSS) is different. Computed tomography (CT) is helpful in distinguishing degenerative and developmental LSS due to its advantage in imaging of osseous and calcified tissues. However, boundaries of the vertebral body, spinal canal and dural sac have low contrast and hard to identify in a CT image, so the diagnosis depends heavily on the knowledge of expert surgeons and radiologists. In this paper, we develop an automatic lumbar spinal CT image segmentation method to assist LSS diagnosis. The main contributions of this paper are the following: 1) a new lumbar spinal CT image dataset is constructed that contains 2393 axial CT images collected from 279 patients, with the ground truth of pixel-level segmentation labels; 2) a dual densely connected U-shaped neural network (DDU-Net) is used to segment the spinal canal, dural sac and vertebral body in an end-to-end manner; 3) DDU-Net is capable of segmenting tissues with large scale-variant, inconspicuous edges (e.g., spinal canal) and extremely small size (e.g., dural sac); and 4) DDU-Net is practical, requiring no image preprocessing such as contrast enhancement, registration and denoising, and the running time reaches 12 FPS. In the experiment, we achieve state-of-the-art performance on the lumbar spinal image segmentation task. We expect that the technique will increase both radiology workflow efficiency and the perceived value of radiology reports for referring clinicians and patients.
\end{abstract}


\titlepgskip=-15pt

\maketitle

\section{Introduction}
\label{sec1}
Lumbar spinal stenosis (LSS) is one of the most common diseases encountered in spinal surgery practice. Diagnosis of LSS is usually made under the guidance of medical imaging techniques such as magnetic resonance imaging (MRI) and computed tomography (CT). More previous studies prefer MRI because it is safer and does not involve any radiation. However, the pathogenesis of degenerative LSS and developmental LSS differ \cite{kitab2018redefining}. Degeneration of the lumbar intervertebral disc, hypertrophy of the articular process and the calcification of ligamentum flavum are the main causes of degenerative LSS; treatment for patients is usually lumbar decompression. Developmental LSS is usually due to vertebral laminae osseous stenosis, and the corresponding treatment is usually laminectomy. Precisely identifying the vertebral body, spinal canal and dural sac is helpful in diagnosing different types of LSS \cite{kuo2019degenerative}. Surgeons usually use lumbar spinal CT images to distinguish between degenerative and developmental LSS because CT is better at imaging osseous and calcified tissues than MRI is \cite{eisenstein1983lumbar}. However, boundaries of the spinal canal and dural sac in CT images are not intuitive; segmentation of these two tissues depends heavily on expert surgeons and radiologists, which brings uncertainty and risk. In this paper, we provide a sufficiently labeled lumbar spinal CT image dataset; the areas of spinal canal, dural sac and vertebral body are labeled in pixel-level. We hope this new dataset will promote the automatic diagnosis of LSS. We then propose a multi-scale densely connected neural network that can automatically segment the spinal canal, dural sac and vertebral body from a raw CT image. To the best of our knowledge, this is the first deep learning-based method to simultaneously segment the spinal canal, dural sac and vertebral body from CT images.

Recently, deep convolutional neural networks have been applied in medical image analysis for it providing abundant and discriminative image representations. Feng et al. \cite{feng2019ccnet} segmented retinal vessels by a cross-connected convolutional network and multi-scale features. Baldeon-Calisto et al. \cite{baldeon2019adaresu} segmented medical images by a multiobjective adaptive convolutional neural network. Li et al. \cite{li2019deep} proposed a 3D fully convolutional network to rationally fuse the complementary information in PET/CT for accurate tumor segmentation. Han et al. \cite{han2014liss} introduced a lung CT imaging signs dataset and proposed a software of abnormal regions annotation. Yu et al. \cite{yu2018melanoma} presented a melanoma recognition method by both a deep learning method and a local descriptor encoding strategy. Nie et al. \cite{nie2018medical} used deep convolutional adversarial networks to synthesize more medical images. Abbati et al. \cite{abbati2017mri} proposed a automatical treatment decision-making plan for LSS. Chen et al. \cite{chen2019brain} used a deep convolutional symmetric neural network to segment brain tumors. In practice, medical image collection and labeling is expensive and time-consuming, however, training deep neural networks usually requires a massive number of training samples. In this paper, we perform data augmentation of the CT images to overcome this limitation. Moreover, we include several dense blocks \cite{huang2017densely} in the proposed dual densely connected U-shaped network (DDU-Net) to reduce the number of parameters and increase the computation efficiency. These two attempts will alleviate the gradient vanishing problem when training a deep neural network with limited data and will improve prediction accuracy as well. In the experiment, we find that some dim-small tissues (e.g., dural sac) are difficult to segment from the original CT image, and the scales of tissues in CT images show large variance. To handle these problems, the proposed DDU-Net contains two U-shaped sub-networks with different sizes of receptive filed, which allows DDU-Net to perform extraction of multi-scale features and segmentation of different sizes of tissues automatically and precisely.

In the experimental section, we test our method on the proposed new dataset and compare the performance with three state-of-the-art image segmentation methods, i.e., U-Net \cite{ronneberger2015u}, FCN \cite{long2015fully} and DeepLab \cite{chen2017rethinking}. Both visual comparison and quantitative comparison show that our method outperforms these state-of-the-art methods.

In summary, this paper makes the following contributions:
\begin{enumerate}
  \item A new challenging dataset is collected for further research and evaluation of spinal CT image segmentation;
  \item Unlike previous works that produce a binary segmentation, this is the first work to segment the spinal canal, dural sac and vertebral body from a spinal CT image simultaneously; we hope that this work will promote automatic diagnosis of lumbar spinal stenosis;
  \item The proposed DDU-Net segment spinal CT images in an automatic and end-to-end manner; all parameters are optimized simultaneously. DDU-Net is capable of segmenting tissues with scale-variant, inconspicuous edges (e.g., spinal canal) and extremely small size (e.g., dural sac);
  \item The proposed method is practical; it requires no image pre-processing such as image registration, denoising, or contrast enhancement. The proposed DDU-Net has only 54M parameters, and it outperforms state-of-the-art methods in both visual comparison and quantitative comparison, with the running time reaching 12 FPS.
\end{enumerate}

The rest of this paper is organized as follows. In section \ref{sec2}, we introduce some previous papers that are related to our work, including medical image analysis and basic deep learning technology. In section \ref{sec3}, the new lumbar spinal CT image dataset is demonstrated. Section \ref{sec4} covers the methodology part of this paper, where we introduce the data augmentation method and architecture of DDU-Net, explaining the details of the network training. In Section \ref{sec5}, visual comparison, qualitative and quantitative comparison are conducted on the proposed method and state-of-the-art methods. Finally, the conclusion is described in Section \ref{sec6}.

\section{Related works}
\label{sec2}
Several state-of-the-art methods for spinal image segmentation have been developed over the past ten years. Some methods have used traditional machine learning and image processing technology, for example, \cite{de2015automatic} and \cite{de2017sct} developed an automatic method for spinal cord and spinal canal segmentation for CT images. Their method is based on multi-resolution propagation of tubular deformable models, and coupled with an automatic intervertebral disk identification method.

With the remarkable performance of deep convolutional neural networks (DCNNs) in different domains such as natural image classification \cite{krizhevsky2012imagenet}, \cite{simonyan2014very} and segmentation \cite{long2015fully,chen2017rethinking}, biomedical image segmentation has achieved a breakthrough by using a U-shaped fully convolutional network (FCN). U-Net \cite{ronneberger2015u} is an end-to-end architecture used to segment different semantics of images, owing to skip connections, this method won the ISBI cell tracking challenge 2015 by using only 30 training images, outperforming the second best method by a large margin. Since then, deep convolutional networks have become popular in automatic biomedical image segmentation. Korez et al. proposed an automatic method to segment vertebral bodies from 3D MRI images. Abbati et al. \cite{abbati2017mri} introduced MRI-based surgical planning for lumbar spinal stenosis, developing an automated algorithm to localize the stenosis causing the patient's symptoms from the MR image; before training the network, the authors manually cropped the original images to obtain the region of interest and trained the network with both labeled and unlabeled images, and the results demonstrated promising performance. Korez et al. \cite{korez2016model} segmented vertebral bodies from MR images with 3D CNNs. Gros et al. \cite{gros2019automatic} segmented both spinal cord and intramedullary multiple sclerosis lesions by convolutional neural networks (CNNs).

In contrast with the aforementioned works, in this paper, we introduce a new fully convolutional network to segment the spinal canal, dural sac and vertebral body in parallel. The proposed method is automatic and does not require any image pre-processing, and the performance surpasses that of state-of-the-art methods.

\section{A new dataset}
\label{sec3}
As this is the first attempt to simultaneously segment the spinal canal, dural sac and vertebral body from CT images, to promote the study of this problem, we have built a new dataset with pixel-level labels. We collected 2393 axial lumbar spinal CT images from 279 patients.

We consider lumbar spinal image segmentation as a pixel-level multi-class classification task, where the input is CT images from different patients of different views and the expected output is a mask with 4 classes, i.e., spinal canal, dural sac, vertebral body and background. Since the spinal canal, dural sac and vertebral body are unique in each CT image, we treat this pixel-level multi-class classification task as a semantic segmentation problem. To ensure label consistency, we asked four radiologists to label four different semantics in all images using a custom designed interactive segmentation tool. We only kept the images that were given very similar labels by all four radiologists. Finally, the proposed dataset contains 1280 images with precise and consistent pixel-level labels. We randomly divided the dataset into three parts, i.e., 50\% for training, 20\% for validation and 30\% for testing.

The first column of Fig. \ref{fig:1} shows 2 sample CT images from our dataset, each image shows a CT scan acquired from an individual patient. The second column is the corresponding ground truth of each raw image. Each ground truth has the same size as the raw image; the red mask of a ground truth indicates the vertebral body region, the green mask of a ground truth indicates the spinal canal region, and the white mask of a ground truth indicates dural sac region. These masks are unique in one ground truth.  Raw images in our dataset have obvious variations, e.g., scale, rotation, brightness, and noise. As shown in Fig. \ref{fig:1}, the scale of the top-left of image Fig. 1(a) is smaller than that of the bottom-left image Fig. 1(b), that is, the details of Fig. \ref{fig:1}(a) are more abundant, and we can see more tissue in Fig. \ref{fig:1}(b). Furthermore, Fig. \ref{fig:1}(a) is more noisy than Fig. \ref{fig:1}(b), a vertical line exists throughout Fig. \ref{fig:1}(a), and Fig. \ref{fig:1}(b) is slightly rotated from the standard visual angle shown in Fig. \ref{fig:1}(a). On the other hand, as shown in the enlarged view of Fig. \ref{fig:1}(b), the boundaries of the spinal canal and dural sac have low contrast against the nearby regions, and the size of the dural sac is extremely small, making it difficult to identify in practical CT images. In sum, the variations and low contrast of the raw CT images make lumbar spinal CT image segmentation difficult, and this dataset is challenging. In this paper, we will develop a robust method to segment the raw CT images in an end-to-end manner, without image denoising, contrast enhancement, registration, etc.

\Figure[t!](topskip=0pt, botskip=0pt, midskip=0pt)[width=10 cm]{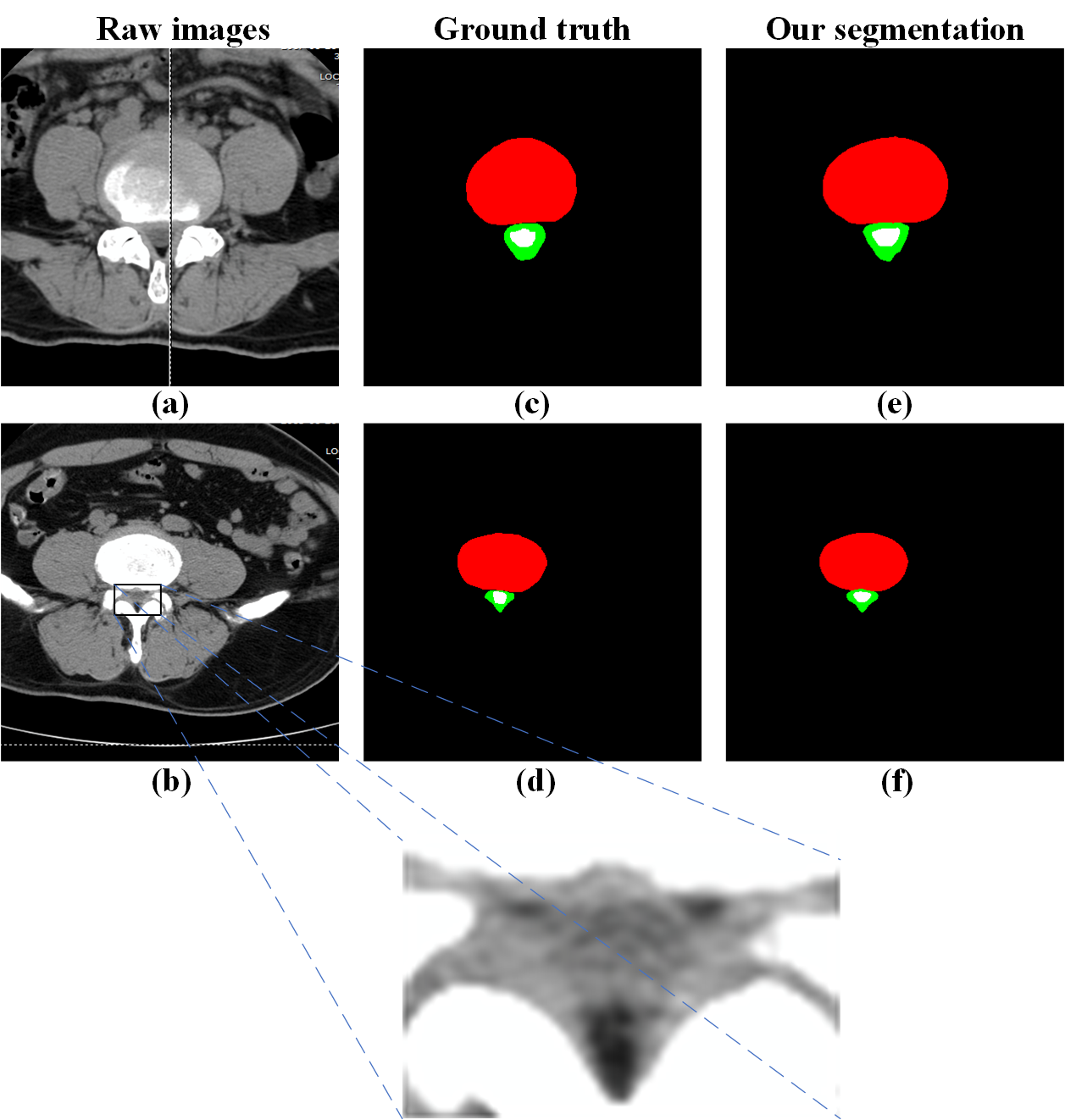}
{Examples of the proposed dataset, from the left to right columns, are raw CT images, ground truth label and our segmentation. The bottommost image shows an enlarged view of the spinal canal and dural sac regions in Fig. 1(b). The red regions denote the vertebral body, the green regions denote the spinal canal, the white regions inside green regions denote the dural sac, and the black regions denote background. \label{fig:1}}

\textbf{Sample imbalance of the dataset.} The labels in our dataset are imbalanced; see Fig. \ref{fig:2} for the class distribution: background (black color) 95.02\%, vertebral body (red color) 4.43\%, spinal canal (green color) 0.37\%, and dural sac (white color) 0.18\%. We will handle this problem by a weighted cross-entropy loss function, please see Sec. 4.3 for details.

\Figure[t!](topskip=0pt, botskip=0pt, midskip=0pt)[width=7.5 cm]{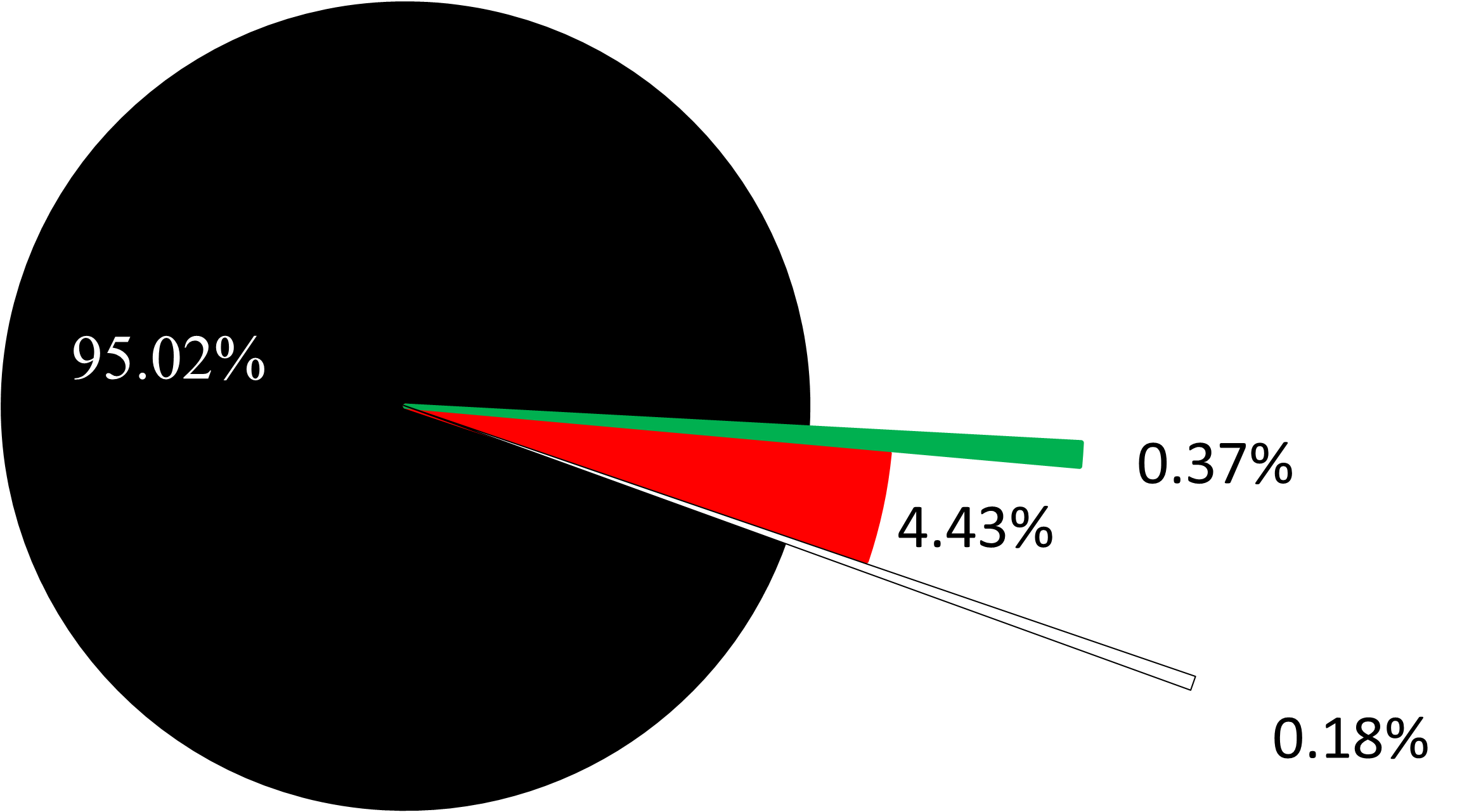}
{Class distribution in the proposed dataset. Black, background; red, vertebral body; green, spinal canal; white, dural sac. This class distribution shows heavily sample imbalance. \label{fig:2}}

\section{Method}
\label{sec4}
In this section, we will introduce details regarding how we segment spinal images automatically and precisely with limited data and why this method works. First, we augment the image using several image processing approaches. This data augmentation alleviates overfitting when training, and we report the performance comparison with and without data augmentation in an ablation study. Second, we construct a dual densely connected U-shaped network (DDU-Net) to segment the spinal canal, dural sac and vertebral body in parallel. Finally, we introduce how to train this network in detail.

\subsection{Data augmentation}
For convolutional neuron network training, we use the following data augmentation: rotation by a random angle between (0, 2$\pi$); horizontal flip of the original images; random crop of the images to a size of $400 \times 400$ from the original size of $512 \times 512$; and random standard Gaussian noise is applied to the images, with the standard deviation $\sigma=0.15+1.15\times random()$, where the random function produces a random float value between 0 and 1. Each image is augmented 100 times by the methods, which alleviates the requirement of a large quantity of labeled data and allow us to train the convolutional neural network successfully.

\subsection{DDU-Net architecture}

\Figure[t!](topskip=0pt, botskip=0pt, midskip=0pt)[width=12 cm]{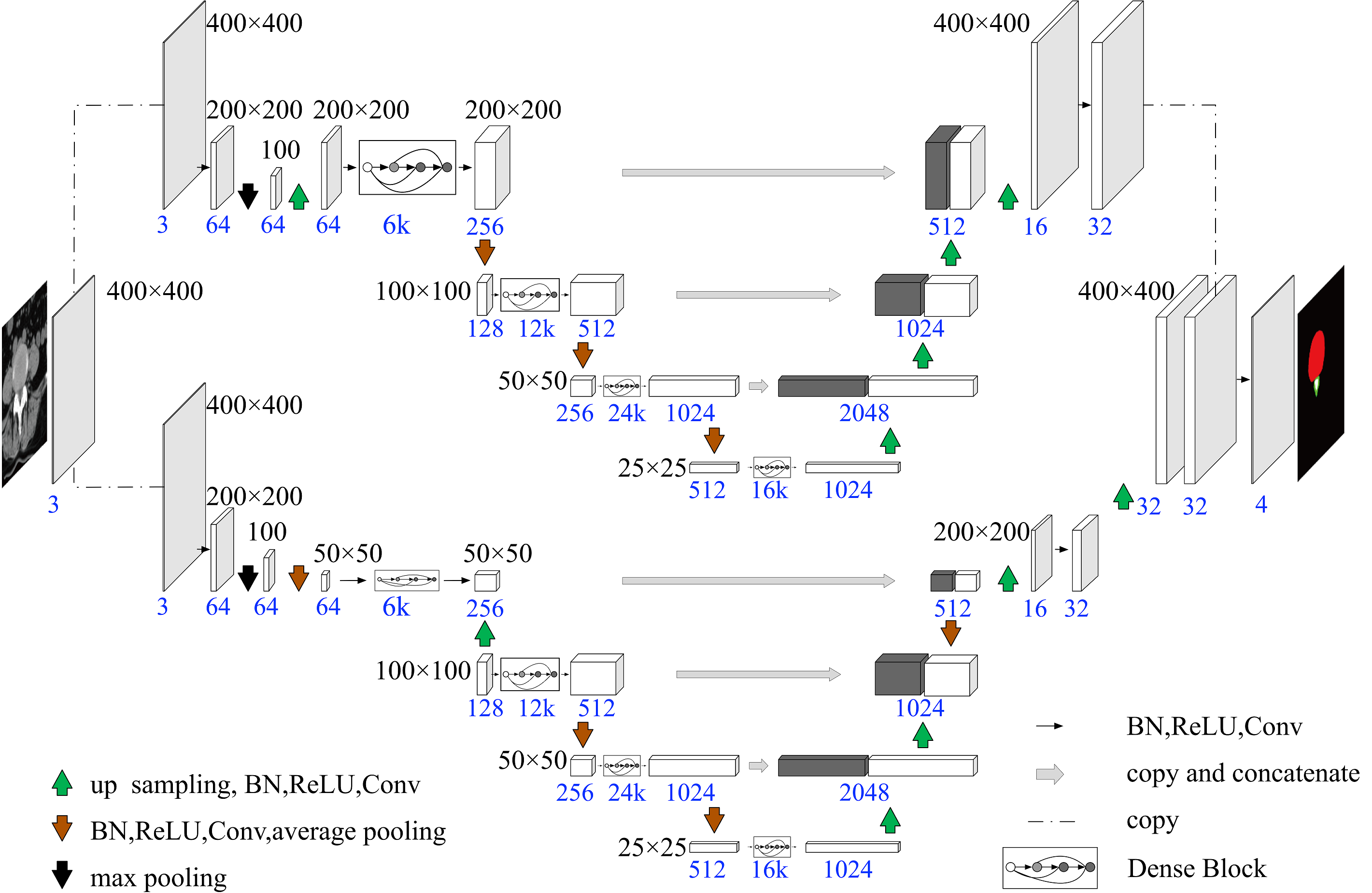}
{DDU-Net architecture, example for input image with size $400\times400$. Each cuboid corresponds to a multi-channel feature map. The grey cuboids represent copied feature maps from prior layers. Dense blocks are built on the downsampling part of the network. The arrows denote the different operations. The color version provides a better view. \label{fig:3}}

We propose a deep fully convolutional network to segment the CT images. Different from U-Net \cite{ronneberger2015u} and some related medical image segmentation methods like \cite{wu2018multiscale} and \cite{jin2019dunet}, the proposed method does not split the original large images into patches, the input of the proposed DDU-Net is the original large images with size $400\times400$, which avoids the need to recompose the image patches. The network architecture is shown in Fig. \ref{fig:3}; the cuboids represent feature maps, and the grey cuboids are copied feature maps from prior layers. In Fig. \ref{fig:3}, the arrows are connections between layers, where solid thin arrows represent standard batch normalization (BN) -rectified linear unit (ReLU)-convolution (Conv); green arrows represent upsampling-BN-ReLU-Conv, and the size of a feature map will increase after this operation; brown arrows represent BN-ReLU-Conv-average pooling, and the size of a feature map will decrease after this operation; black arrows represent max pooling, and the size of a feature map will also decrease after this operation; and grey arrows represent copying the source feature maps and concatenating them to the target feature maps, and the channels of the target layer will increase after this operation. Black numbers are the size of the feature maps, and blue numbers are the channels of the layers.

\textbf{Dual network structure.} DDU-Net consists of two sub-networks, and we duplicate an image and feed it into the two sub-networks separately. The upper sub-network upsamples feature maps at the fourth layer and downsamples at the fifth layer, while the lower sub-network downsamples feature maps at the fourth layer and upsamples at the fifth layer. Neurons of the upper sub-network have a smaller receptive field than the lower sub-network; consequently, the upper sub-network concentrates smaller tissues, and the lower sub-network concentrates larger tissues. We merge the feature maps of the last layer from the two sub-networks and convolve them with a $1\times1\times4$ convolution layer to obtain a pixel-level classification. This dual network architecture makes DDU-Net segment different size of tissues robustly, e.g., Fig. \ref{fig:1}(c) and Fig. \ref{fig:1}(d) shows the small dural sac (white color) and the large vertebral body (red color).

\textbf{Skip connections.} Inspired by U-Net \cite{ronneberger2015u}, DDU-Net consists of a downsampling part (left side) and an upsampling part (right side). The downsampling part is used to encode input images in a lower dimensionality with richer filters, while the upsampling part is designed to complete the inverse process of encoding by upsampling and merging low-dimensional feature maps, which produce dense predictions of each pixel. On each sub-network, skip connections copy feature maps from the $5^{th}$, $7^{th}$ and $9^{th}$ layers, and concatenate the feature maps to the $12^{th}$, $13^{th}$ and $14^{th}$ layers, respectively; as shown in Fig. \ref{fig:3} grey arrows indicate copy directions, and grey cuboids represent duplicated feature maps.

\textbf{Dense blocks.} Generally, a deeper network performs better than a shallower network. To balance the depth of network and number of parameters, we insert dense blocks at the downsampling part of DDU-Net. Inside the dense blocks, neurons of each layer connect not only to the next layer but also to all their subsequent layers:
\begin{equation}
L_{n}=\Gamma_{n} \left ( \left [ L_{1},L_{2},...,L_{n-1} \right ] \right ),
\label{eq:1}
\end{equation}
where $L_{n}$  is the $n^{th}$ layer and $\Gamma_{n}()$ is a transition function, which is usually BN-ReLU-Conv 3 consecutive operations, and   affected by $L_{1},L_{2},...,L_{n-1}$.

Fig. \ref{fig:4} shows an illustration of a dense block; between the input and output, there are 3 layers, with each layer having $k=2$ channels, and we call it a $3k$ dense block. In this example, layer 1 connects to layer 2, as well as to layer 3 and the output layer. As shown in Fig. \ref{fig:3}, we designed 8 dense blocks in DDU-Net in total, i.e., $6k, 12k, 24k, $ and $16k$ dense blocks for each sub-network.

\Figure[t!](topskip=0pt, botskip=0pt, midskip=0pt)[width=10 cm]{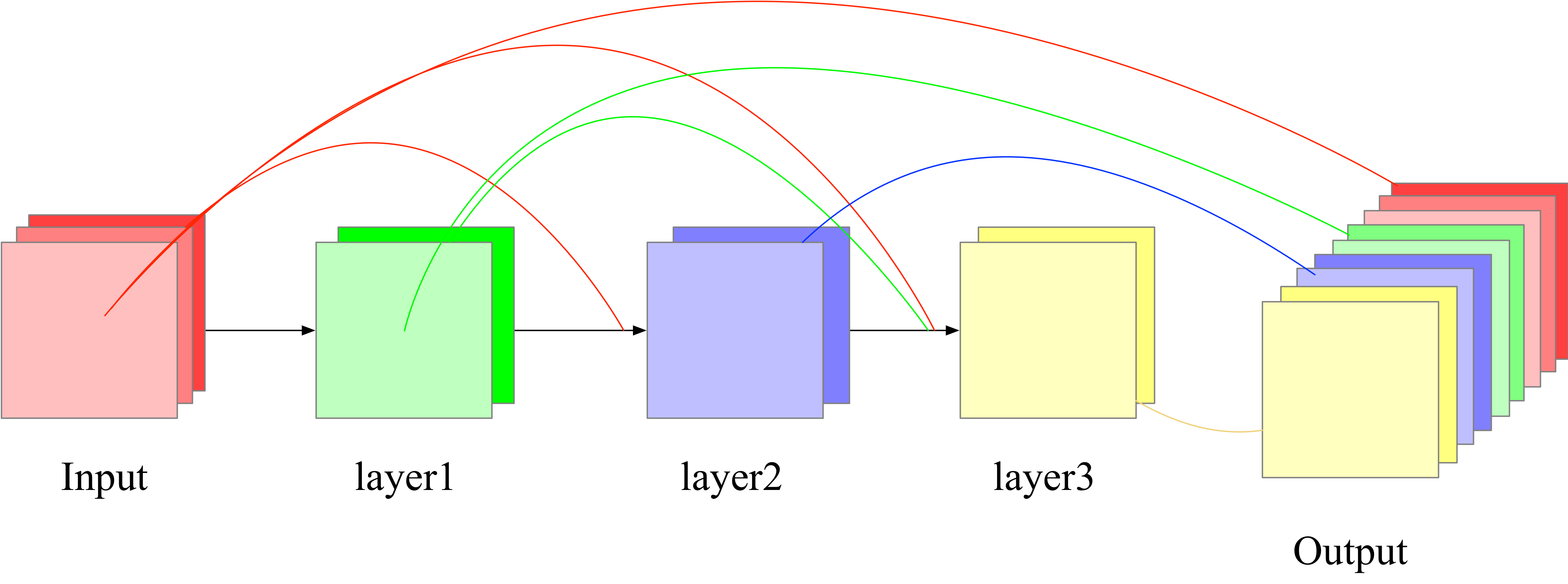}
{A 3-layers dense block with growth rate $k=2$. Each layer has 2 channels, and each layer connect not only to the next layer but also to all their subsequent layers. \label{fig:4}}

\textbf{Detailed structure of DDU-Net.} See Table \ref{tab:1} for a detailed structure of the upper sub-network and lower sub-network of DDU-Net. Please note that in the layer details, conv is BN-ReLU-Conv; $7 \times 7$, 64 conv, stride 2 corresponds to the sequence BN-ReLU-Conv layer with convolutional kernel size of $7 \times 7$ and 64 channels with stride 2; the symbol $[] \times n$ represent a dense block with operations in $[]$ repeating $n$ times, and $-[]$ means that this layer skip-connects with a dense block. The growth rate for all dense blocks is $k=32$; the upsampling is bilinear interpolation, and each transition layer is an operation between two dense blocks. The lower sub-network has one additional upsamplinig layer to recover the feature map size to $400\times400$. The results of the upper sub-network and lower sub-network are concatenated and convolved with a $1\times1\times4$ convolutional layer for dense classification.

\begin{table*}[]
\caption{The architecture of the two sub-networks of DDU-Net.}
\scriptsize
\centering
\begin{tabular}{|c|c|c|c|c|c|}
\multicolumn{3}{c}{\textbf{Upper sub-network}}                                                                                                                         & \multicolumn{3}{c}{\textbf{Lower sub-network}}                                                                                                                                                \\ \hline
Layer name         & Layer details                                                                           & Feature size                                              & Layer name         & Layer details                                                                                                  & Feature size                                             \\ \hline
input              & -                                                                                       & 400$\times$400                                                   & input              & -                                                                                                              & 400$\times$400                                                  \\ \hline
Convolution 1      & 7$\times$7, 64 conv, stride 2                                                                   & 200$\times$200                                                   & Convolution 1      & 7$\times$7, 64 conv, stride 2                                                                                         & 200$\times$200                                                  \\ \hline
Pooling            & 3$\times$3, 64 max pool, stride 2                                                              & 100$\times$100                                                   & Pooling            & 3$\times$3, 64 max pool, stride 2                                                                                     & 100$\times$100                                                  \\ \hline
Upsampling 1       & 2$\times$2, 64 upsampling                                                                      & 200$\times$200                                                   & Transition Layer 1 & 2$\times$2, 64 average pool, stride 2                                                                               & 50$\times$50                                                    \\ \hline
Dense Block 1      & $\begin{bmatrix} 1\times1, 128\\ 3\times3, 32 \end{bmatrix} \times 6$                                                 & 200$\times$200                                                   & Dense Block 1      & $\begin{bmatrix} 1\times1, 128\\ 3\times3, 32 \end{bmatrix} \times 6$                                           & 50$\times$50                                                    \\ \hline
Transition Layer 1 & \begin{tabular}[c]{@{}c@{}}1$\times$1, 128 conv\\ 2$\times$2, 128 average pool, stride 2\end{tabular} & \begin{tabular}[c]{@{}c@{}}200$\times$200\\ 100$\times$100\end{tabular} & Upsampling 1       & 2$\times$2, 128 upsampling                                                                                          & 100$\times$100                                                  \\ \hline
Dense Block 2      & $\begin{bmatrix} 1\times1, 128\\ 3\times3, 32 \end{bmatrix} \times 12$                   & 100$\times$100                                                   & Dense Block 2      & $\begin{bmatrix} 1\times1, 128\\ 3\times3, 32 \end{bmatrix} \times 12$                                          & 100$\times$100                                                  \\ \hline
Transition Layer 2 & \begin{tabular}[c]{@{}c@{}}1$\times$1, 256 conv\\ 2$\times$2, 256 average pool, stride 2\end{tabular} & \begin{tabular}[c]{@{}c@{}}100$\times$100\\ 50$\times$50\end{tabular}   & Transition Layer 2 & \begin{tabular}[c]{@{}c@{}}1$\times$1, 256 conv\\ 2$\times$2, 256 average pool, stride 2\end{tabular}                          & \begin{tabular}[c]{@{}c@{}}100$\times$100\\ 50$\times$50\end{tabular}  \\ \hline
Dense Block 3      & $\begin{bmatrix} 1\times1, 128\\ 3\times3, 32 \end{bmatrix} \times 24$                  & 50$\times$50                                                     & Dense Block 3      & $\begin{bmatrix} 1\times1, 128\\ 3\times3, 32 \end{bmatrix} \times 24$                                         & 50$\times$50                                                  \\ \hline
Transition Layer 3 & \begin{tabular}[c]{@{}c@{}}1$\times$1, 512 conv\\ 2$\times$2, 512 average pool, stride 2\end{tabular} & \begin{tabular}[c]{@{}c@{}}50$\times$50\\ 25$\times$25\end{tabular}     & Transition Layer 3 & \begin{tabular}[c]{@{}c@{}}1$\times$1, 512 conv\\ 2$\times$2, 512 average pool, stride 2\end{tabular}                        & \begin{tabular}[c]{@{}c@{}}50$\times$50\\ 25$\times$25\end{tabular}    \\ \hline
Dense Block 4      & $\begin{bmatrix} 1\times1, 128\\ 3\times3, 32 \end{bmatrix} \times 16$                  & 25$\times$25                                                     & Dense Block 4      & $\begin{bmatrix} 1\times1, 128\\ 3\times3, 32 \end{bmatrix} \times 16$                                         & 25$\times$25                                                    \\ \hline
Upsampling 2       & 2$\times$2 upsampling - {[}Dense Block3{]}, 2048                                             & 50$\times$50                                                     & Upsampling 2       & 2$\times$2 upsampling - {[}Dense Block3{]}, 2048                                                                      & 50$\times$50                                                    \\ \hline
Convolution 2      & \begin{tabular}[c]{@{}c@{}}1$\times$1, 512 conv\\ 3$\times$3, 512 conv\end{tabular}                   & 50$\times$50                                                     & Convolution 2      & \begin{tabular}[c]{@{}c@{}}1$\times$1, 512 conv\\ 3$\times$3, 512 conv\end{tabular}                                          & 50$\times$50                                                    \\ \hline
Upsampling 3       & 2$\times$2 upsampling - {[}Dense Block 2{]}, 1024                                              & 100$\times$100                                                   & Upsampling 3       & 2$\times$2 upsampling - {[}Dense Block 2{]}, 1024                                                                     & 100$\times$100                                                  \\ \hline
Convolution 3      & \begin{tabular}[c]{@{}c@{}}1$\times$1, 256 conv\\ 3$\times$3, 256 conv\end{tabular}                   & 100$\times$100                                                   & Convolution 3      & \begin{tabular}[c]{@{}c@{}}1$\times$1, 256 conv\\ 3$\times$3, 256 conv\end{tabular}                                          & 100$\times$100                                                  \\ \hline
Upsampling 4       & 2$\times$2 upsampling - {[}Dense Block 1{]}, 512                                               & 200$\times$200                                                   & Transition Layer 4 & \begin{tabular}[c]{@{}c@{}}1$\times$1, 256 conv\\ 2$\times$2 average pool, stride 2 - {[}Dense Block 1{]}, 512\end{tabular} & \begin{tabular}[c]{@{}c@{}}100$\times$100\\ 50$\times$50\end{tabular} \\ \hline
Convolution 4      & \begin{tabular}[c]{@{}c@{}}1$\times$1, 16 conv\\ 3$\times$3, 16 conv\end{tabular}                     & 200$\times$200                                                   & Convolution 4      & \begin{tabular}[c]{@{}c@{}}1$\times$1, 16 conv\\ 3$\times$3, 16 conv\end{tabular}                                          & 50$\times$50                                                    \\ \hline
Upsampling 5       & 2$\times$2 upsampling                                                                          & 400$\times$400                                                   & Upsampling 5       & 4$\times$4, 16 upsampling                                                                                           & 200$\times$200                                                 \\ \hline
Convolution 5      & \begin{tabular}[c]{@{}c@{}}1$\times$1, 32 conv\\ 3$\times$3, 32 conv\end{tabular}                     & 400$\times$400                                                   & Convolution 5      & \begin{tabular}[c]{@{}c@{}}1$\times$1, 32 conv\\ 3$\times$3, 32 conv\end{tabular}                                            & 200$\times$200                                                  \\ \hline
-                  & -                                                                                       & -                                                         & Upsampling 6 & 2$\times$2, 32 upsampling                                                                                             & 400$\times$400                                                  \\ \hline
\end{tabular}
\label{tab:1}
\end{table*}

\subsection{Training}
We randomly divide the proposed dataset into three parts, i.e., 50\% for training, 20\% for validation and 30\% for testing. In practice, 50\% of the images are fed into DDU-Net for training, 20\% of the images are used for hyperparameters optimization and prevention of overfitting, and 30\% of the images are used to evaluate the performance of the neural networks. The DDU-Net is trained in an end-to-end manner, and all parameters in the network are optimized simultaneously.

For a typical lumbar spinal CT image, most pixels are the background, and regions such as the dural sac and spinal canal are extremely small (see Fig. \ref{fig:1} and Fig. \ref{fig:2}). To solve this problem, we introduce a class-balancing weight $w^c$ on a per-pixel term basis, this class-balancing weight is designed to offset the imbalance between major and minor classes and promote the neural network to learn features of small tissues such as the dural sac and spinal canal. Specifically, the DDU-Net adopts a weighted cross-entropy function as the loss function, which can be formulated as:
\begin{equation}\label{eq:2}
\mathcal{L}(y,p)=-\frac{1}{N}\sum_{i=1}^{N}\sum_{c=1}^{4} w^c \left [y_i^c logp_i^c+\left (1-y_i^c  \right )log\left (1-p_i^c  \right )  \right ],
\end{equation}
where $N$ is pixel number of the image, $y_i^c$ is the labeled class of the pixel $i$, $p_i^c$is the prediction probability of pixel $i$ belonging to class $c$, i.e., spinal canal ($c=1$), dural sac ($c=2$), vertebral body ($c=3$) and background ($c=4$), and $w^c$ denotes the class-balancing weight as:
\begin{equation}\label{eq:3}
w^c=e^{-\frac{N^c}{N}},
\end{equation}
where $N$ is pixel number of the image and $N^c$ is the pixel number that belongs to class $c$.

The code of DDU-Net is implemented by Pytorch framework. We train DDU-Net using two NVIDIA GeForce GTX 1080Ti GPUs, and due to GPU memory constraints, our model is trained with a mini-batch size 4. The optimizer that we adopt is stochastic gradient descent (mini-batch SGD) \cite{sutskever2013importance} with momentum=0.95; the learning rate is set to 1e-7, and the weight decay is 5e-4. The parameters in dense blocks of each sub-network are initialized with DenseNet \cite{huang2017densely} weights pretrained on ImageNet \cite{russakovsky2015imagenet}, and other parameters are initialized by He initialization \cite{he2015delving}.

\section{Experiments and results}
\label{sec5}
In this section, we introduce the performance evaluation metrics of this paper, after which ablation studies are conducted to explain why this method works. Finally, we compare the DDU-Net with state-of-the-art methods.

\subsection{Evaluation metrics}
Suppose that we classify image pixels into $C$ classes, where $N$ is the total pixel number of the image, $\#y_{ii}$ is the correct predicted pixel number, and $\#y_{ij}$ and $\#y_{ji}$ are false positive and false negative predicted pixel numbers respectively, as shown in Fig. \ref{fig:5}. We evaluated our model using several semantic segmentation metrics \cite{garcia2017review}: pixel accuracy (PA), mean pixel accuracy (MPA), mean intersection over union (MIoU), and frequency weighted intersection over union (FWIoU).

\Figure[t!](topskip=0pt, botskip=0pt, midskip=0pt)[width=7.5 cm]{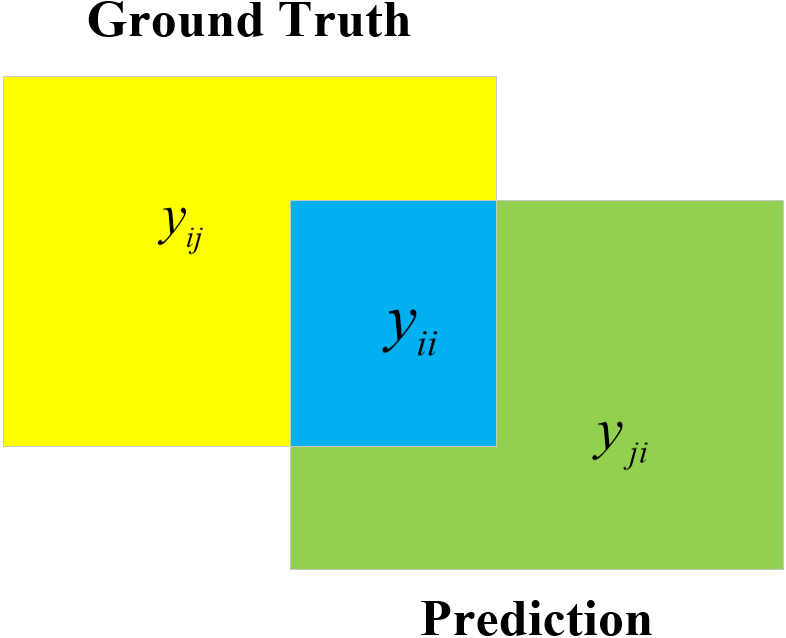}
{An illustration of semantic segmentation: $y_{ii}$ is a correctly predicted pixel, $y_{ij}$ is an i-class pixel that was predicted as j-class, and $y_{ji}$ is a j-class pixel that was predicted as i-class. \label{fig:5}}

PA measures the ratio between correctly predicted pixels and total pixels. This metric can be formulated as follows:
\begin{equation}\label{eq:4}
PA=\frac{\sum_{i=0}^{C}\#y_{ii}}{\sum_{i=0}^{C}\sum_{j=0}^{C}\left (\#y_{ii}+\#y_{ij}  \right )}.
\end{equation}

MPA is a simple improvement of PA that calculates the percentage of correctly predicted pixels of each class and averages the percentages as a result. This metric can be formulated as follows:
\begin{equation}\label{eq:5}
MPA=\frac{1}{C}\sum_{i=1}^{C}\frac{\#y_{ii}}{\sum_{j=1}^{C}\left ( \#y_{ii}+\#y_{ij} \right )}.
\end{equation}

mIoU is a common metric in semantic segmentation that calculates the ratio between the intersection region (true positives) and the union region (true positives, false positives and false negatives), and average the ratios on all classes. This metric can be formulated as follows:
\begin{equation}\label{eq:6}
mIoU=\frac{1}{C}\sum_{i=1}^{C}\frac{\#y_{ii}}{\sum_{j=1}^{C}\left ( \#y_{ii}+\#y_{ij}+\#y_{ji} \right )}.
\end{equation}

fwIoU is an improvement of mIoU, it weights the intersection over union of each class by their occurrence rate. This metric can be formulated as follows:
\begin{equation}\label{eq:7}
\begin{split}
fwIoU=\sum_{i=1}^{C}\frac{\sum_{j=1}^{C}\left (\#y_{ij}+\#y_{ii}  \right )}{N} \times\\
\frac{\#y_{ii}}{\sum_{j=1}^{C}\left ( \#y_{ii}+\#y_{ij}+\#y_{ji} \right )}.
\end{split}
\end{equation}

\subsection{Ablation studies}
To investigate the importance of different options in our method, we conduct an ablation study. Ablation studies include: with/without data augmentation, with/without skip connections, with/without dense blocks, with/without multi-branch networks, and the growth rate affect. Table \ref{tab:2} presents the details of the ablation studies, the last row shows the default DDU-Net as the baseline, that is, the network adopts skip connections, dense blocks and multi-branches, and use data augmentation.

\textbf{Data augmentation.} Since the proposed dataset is not a large-scale dataset, we conduct data augmentation to alleviate overfitting when training the network; the data augmentation details are introduced in Section 4.1. To reveal the benefit of data augmentation, we train this model for 100 epochs with and without data augmentation. The last row and the second row of Table \ref{tab:2} show the performance comparison between training with and without data augmentation under the same network architecture; the network adopts skip-connections, dense blocks and multi-branch structure, and after data agumentation, the mIoU improves by approximately 1 point . Fig. \ref{fig:6} shows the training and validation losses during the training procedure with and without data augmentation; the blue curves denote training loss, and the orange curves denote validation loss. Fig. \ref{fig:6}(a) shows the training process without data augmentation, and the best mIoU is $0.8209$ at about the $90^{th}$ epoch. The validation loss begins to increase again (after a fall) at about the $30^{th}$ epoch, this phenomenon is caused by overfitting. Fig. \ref{fig:6}(b) depicts the training process with data augmentation, although the validation curve also increases again at about the $45^{th}$ epoch, with the best mIoU reaching $0.8314$ at about the $90^{th}$ epoch, and improving by approximately 1 point. The above mentioned experiments indicate that data augmentation in our method not only alleviates overfitting but also improves the performance.

\Figure[t!](topskip=0pt, botskip=0pt, midskip=0pt)[width=12 cm]{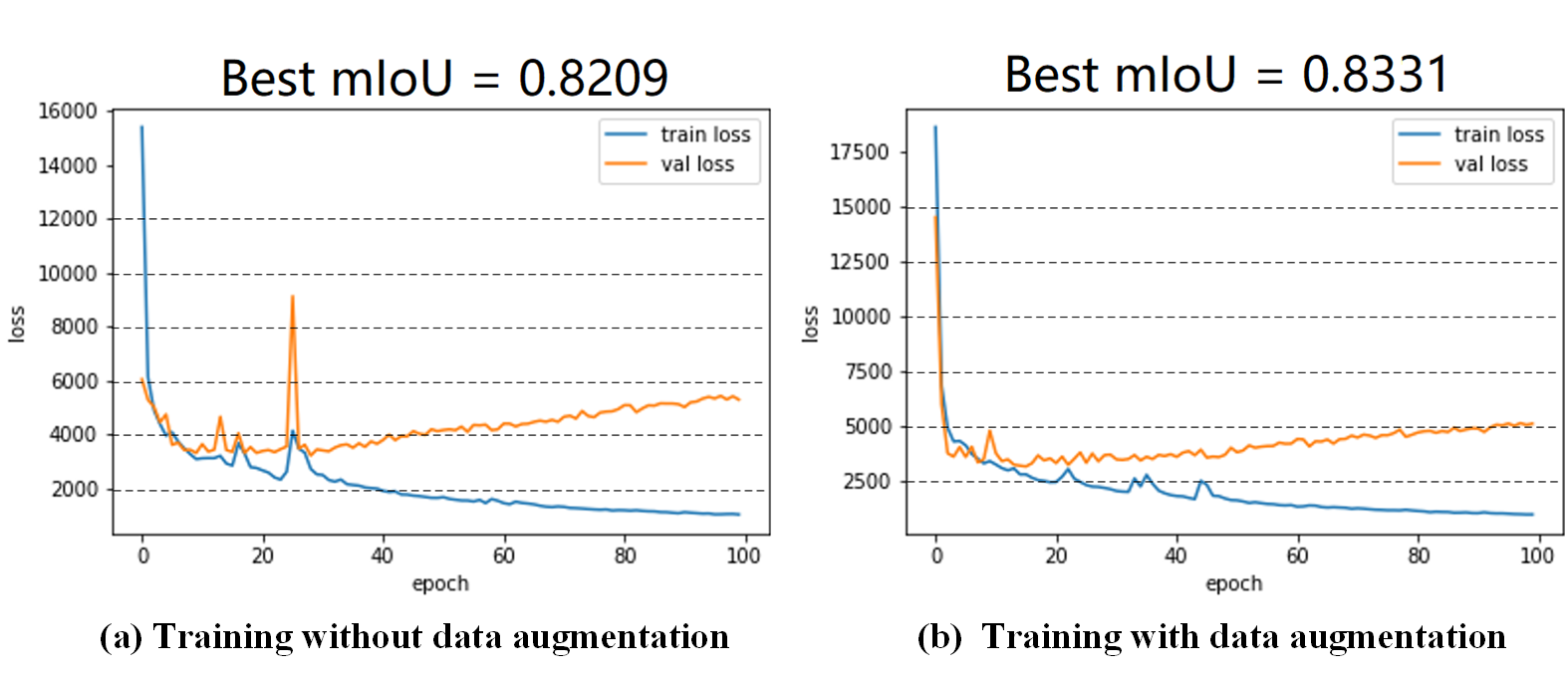}
{Training on the proposed dataset, (a) without data augmentation, (b) with data augmentation. We observe that training with data augmentation alleviates overfitting and improves the performance. \label{fig:6}}

\textbf{Network architecture.}
We design several modules to improve architecture of DDU-Net. Skip connections in the DDU-Net are designed to reuse low-level features and fuse multi-level features. As shown in the third row of Table \ref{tab:2}, we remove all skip connections in DDU-Net, after which the network becomes a \emph{flat network}; the mIoU of the \emph{flat network} is $0.7989$, which is lower than that of the default DDU-Net by 3.42 points. Dense blocks are capable of alleviating the gradient vanishing problem in deep networks, enhancing feature resuse and feature propagation, and also reducing the number of parameters. As shown in Table \ref{tab:2}, without dense blocks, the mIoU decreased to $0.7636$, which is lower than that of the defualt DDU-Net by 6.95 points. We separate DDU-Net into two types: one retains only the upper sub-network, and the other retains only the lower sub-network. The fifth and sixth rows of Table \ref{tab:2} show the performances of stand-alone upper sub-network and lower sub-network, respectively; their mIoU are $0.8186$ and $0.8067$, which are lower than the mIoU achieved by the default DDU-Net consisting of multi-branch networks by 1.45 and 2.64 points. The evaluation results under other metrics such as PA, MPA and fwIoU also indicate that the default DDU-Net which using data augmentation and three modules (SC, DB and MB) performs best.

\begin{table*}[]
\caption{Ablation Study under different options. Four options of the proposed method: data augmentation (DA), skip-connections (SC), dense blocks (DB), and multi-branches (MB).}
\centering
\footnotesize
\begin{tabular}{|l|l|l|l|l|l|l|l|l|}
\hline
DA & SC & DB & MB & mIoU & PA & MPA & fwIoU & Destription  \\\hline
   &\checkmark    &\checkmark    &\checkmark    &0.8209   &0.9909 &0.8901 &0.9828 &Without data augmentation  \\
\checkmark   &    &\checkmark    &\checkmark    &0.7989   &0.9908 &0.8836 &0.9827 &A flat network  \\
\checkmark   &\checkmark    &    &\checkmark    &0.7636   &0.9856 &0.8599 &0.9731 &Without dense blocks   \\
\checkmark   &\checkmark    &\checkmark    &    &0.8186   &0.9905 &0.8862 &0.9821 &Upper sub-network   \\
\checkmark   &\checkmark    &\checkmark    &    &0.8067   &0.9898 &0.8858 &0.9807 &Lower sub-network   \\
\checkmark   &\checkmark    &\checkmark    &\checkmark    &\textbf{0.8331} &\textbf{0.9913} &\textbf{0.9099} &\textbf{0.9835} &DDU-Net (default)  \\
\hline
\end{tabular}
\label{tab:2}
\end{table*}

We visualize the ablation studies in Fig. \ref{fig:7}, where the images in the leftmost column are two input CT images; scale of the top-left image is larger than that of the bottom-right image. The images on rightmost side are the corresponding ground truth (GT), of which the spinal canal, dural sac, vertebral body and background are marked in green, white, red and black, respectively. The columns from Fig. \ref{fig:7}(b) to Fig. \ref{fig:7}(f) show the network predictions under several options, that is, Fig. \ref{fig:7}(b) presents results from the flat network without skip connections, Fig. \ref{fig:7}(c) represents results from the network without dense blocks, Fig. \ref{fig:7}(d) represents results from the upper sub-network, Fig. \ref{fig:7}(e) represents results from the lower sub-network, and Fig. \ref{fig:7}(e) represents the complete DDU-Net segmentations. We can clearly see that the segmentation results generated by the complete DDU-Net are much closer to the ground truth than the other results.

\Figure[t!](topskip=0pt, botskip=0pt, midskip=0pt)[width=15 cm]{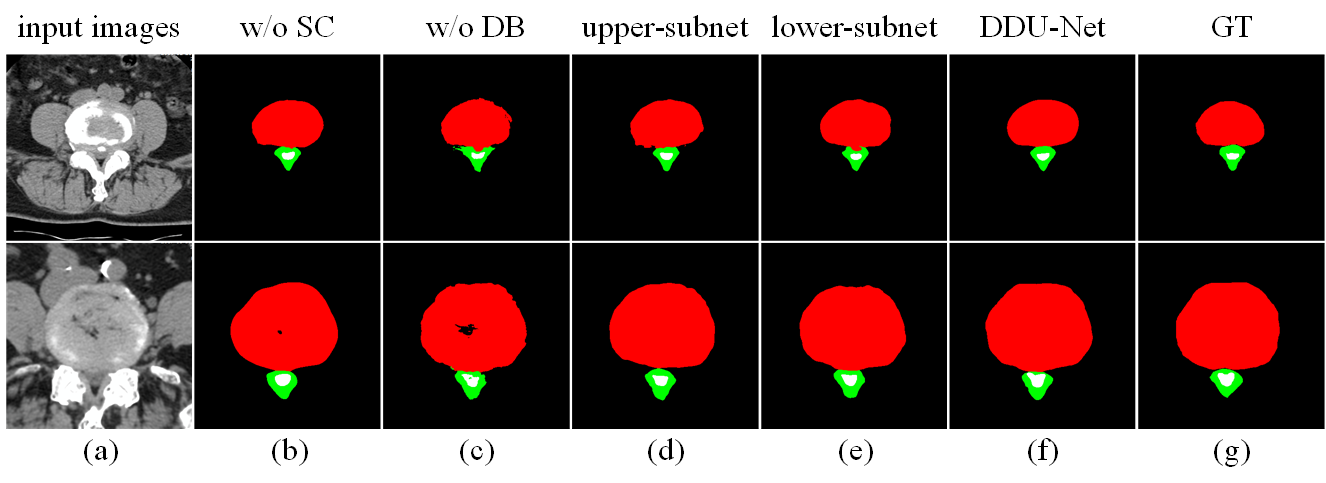}
{Examples from the ablation study. This includes a comparison between the ground truth (rightmost column) and results from different options, which are shown respectively from (b) to (f): w/o SC (without skip connection), w/o DB (without dense blocks), upper-subnet (only upper-subnet), lower-subnet (only lower-subnet), DDU-Net (the complete DDU-Net). \label{fig:7}}

\textbf{Growth rate.}
The growth rate is a hyperparameter of a densely connected neural network that indicates how many layers a dense block has. Generally, a network with a larger growth rate will perform better. However, a larger growth rate will bring about more parameters, and the running time of the network will also increase. In the experiment we test several growth rate numbers under the same network architecture and data. As shown in Table \ref{tab:3}, the method performs best when the growth rate equal to 48, and it surpasses the growth rate 32 by a very small margin. On the other hand, when the growth rate is equal to 32, the running time is much faster than when the growth rate is equal to 48; this is because the former parameter number is much smaller than the latter. Consequently, we set the growth rate equal to 32 as the optimal hyperparameter since it has a good tradeoff between accuracy and efficiency.

\begin{table}[]
\caption{The growth rate impacts parameter the number of parameters and the running time of DDU-Net; we choose a growth rate of $k=32$ as it has a good tradeoff between accuracy and efficiency.}
\centering
\scriptsize
\begin{tabular}{|l|l|l|l|l|l|l|}
\hline
Growth rate $k$ & \#parameters & FPS               & mIoU    & PA     & MPA & fwIoU \\ \hline
12          & 8.62M    & 18.31 & 0.7797 & 0.9873 & 0.8558  & 0.9760 \\
24          & 31.44M   & 16.14 & 0.8183 & 0.9902 & 0.8938  & 0.9815 \\
\textbf{32}          & 54.65M   & 12.35 & 0.8331 & 0.9913 & 0.9099  & 0.9835 \\
48          & 82.21M  & 8.12 & 0.8340 & 0.9934 & 0.9122  & 0.9865 \\ \hline
\end{tabular}
\label{tab:3}
\end{table}

\subsection{Comparison with state-of-the-art methods}
\textbf{Qualitative Comparison}
Several state-of-the-art methods are related to spinal image analysis; since the proposed method is the first to simultaneously segment the vertebral body, spinal canal and dural sac, we compare our DDU-Net with five related state-of-the-art methods in the following three aspects: 1) data source type, either MRI or CT images; 2) technical details, including objective of the work and methodology; and 3) segmentation targets, explaining what contents are segmented from the images. As shown in Table \ref{tab:4}, all methods except our proposed DDU-Net using MR images as training data. \cite{de2015automatic} and \cite{gros2018automatic} segment images by using traditional machine learning algorithms, \cite{abbati2017mri} segment images manually, where the segmented images are fed into CNNs as intermediate results; and other methods segment images by CNNs. The segmentation targets of these methods are different, only the proposed DDU-Net segment the vertebral body, spinal canal and dural sac from spinal CT images.

\begin{table*}[]
\caption{Qualitative comparison of different methods.}
\centering
\footnotesize
\begin{tabular}{|c|c|c|c|}
\hline
Methods & Data source type & Technical details                                                                                                                        & Segmentation targets                                                                       \\ \hline
Leener 2015 \cite{de2015automatic}       & MRI                       & \begin{tabular}[c]{@{}c@{}}Automatically segment spinal cord and\\  spinal canal by vertebral label\end{tabular}                                  & \begin{tabular}[c]{@{}c@{}}Vertebral regions, spinal cord\\ and cerebrospinal fluid\end{tabular}    \\ \hline
Gros 2018 \cite{gros2018automatic}        & MRI                       & \begin{tabular}[c]{@{}c@{}}Automatically localize spinal cord\\ using global curve optimization\end{tabular}                                      & Spinal cord                                                                                         \\ \hline
Gros 2019 \cite{gros2019automatic}       & MRI                       & \begin{tabular}[c]{@{}c@{}}Automatically segment spinal cord and intramedullary\\ multiple sclerosis lesions with CNNs\end{tabular}               & \begin{tabular}[c]{@{}c@{}}Spinal cord and intramedullary\\ multiple sclerosis lesions\end{tabular} \\ \hline
Korez 2016 \cite{korez2016model}      & MRI                       & Automatically segment vertebral body by 3D CNNs                                                                                                   & Vertebral body                                                                                      \\ \hline
Abbati 2017 \cite{abbati2017mri}     & MRI                       & \begin{tabular}[c]{@{}c@{}}Automatically diagnose lumbar spinal stenosis by \\ CNNs, segmentations are intermediate results\end{tabular}          & \begin{tabular}[c]{@{}c@{}}Manual scan chopping and\\ interpolation to four slides\end{tabular}     \\ \hline
DDU-Net (proposed)          & CT images                 & \begin{tabular}[c]{@{}c@{}}Automatically segment vertebral body, spinal canal\\ and dural sac by a dual densely connected U-shaped CNN\end{tabular} & \begin{tabular}[c]{@{}c@{}}Vertebral body, spinal canal\\ and dural sac\end{tabular} \\ \hline
\end{tabular}
\label{tab:4}
\end{table*}

\textbf{Visual Comparison}
For visual comparison, we select some sample segmentation results of three state-of-the-art deep learning-based semantic segmentation methods and DDU-Net. As shown in Fig. \ref{fig:8}, the images in the leftmost column are input images, and the images in the rightmost column are the ground truth. Fig. \ref{fig:8}(b) shows the result for FCN-8s \cite{long2015fully}, which applies per-pixel classification using a fully convolutional network; in this experiment, we adopt the FCN 8 pixel stride version since it performs best among all FCN versions; Fig. \ref{fig:8}(c) shows the results for U-Net \cite{ronneberger2015u}, which segments medical images by a U-shaped fully convolutional network with skip connections; Fig. \ref{fig:8}(d) shows the results for DeeplabV3 \cite{chen2017rethinking}, which improved Deeplab \cite{chen2017deeplab} by multigrid and atrous spatial pyramid pooling; Fig. \ref{fig:8}(e) demonstrates segmentation maps of the proposed DDU-Net. As demonstrated in Fig. \ref{fig:8}, FCN-8s \cite{long2015fully} and U-Net \cite{ronneberger2015u} failed to segment the extremely small dural sac, boundaries of the vertebral body in U-Net \cite{ronneberger2015u} and DeeplabV3 \cite{chen2017rethinking} are not precisely, and we can clearly see that the segmentation maps of DDU-Net are much closer to the GT than those of the other methods. From the top-left to bottom-left image, the scale of the images are increasing, that is, the top-left image concentrates details of tissues, where the regions of interest are larger than those of other images, and the bottom-left image represent a global view of CT, regions of interest are smaller than other images. Under this challenging condition, DDU-Net not only handles images with different scales but also segments semantics with different sizes, such as the large vertebral body (denoted in red) and small dural sac (denoted in white), while FCN-8s, U-Net and DeeplabV3 failed in at least one case.

\Figure[t!](topskip=0pt, botskip=0pt, midskip=0pt)[width=15 cm]{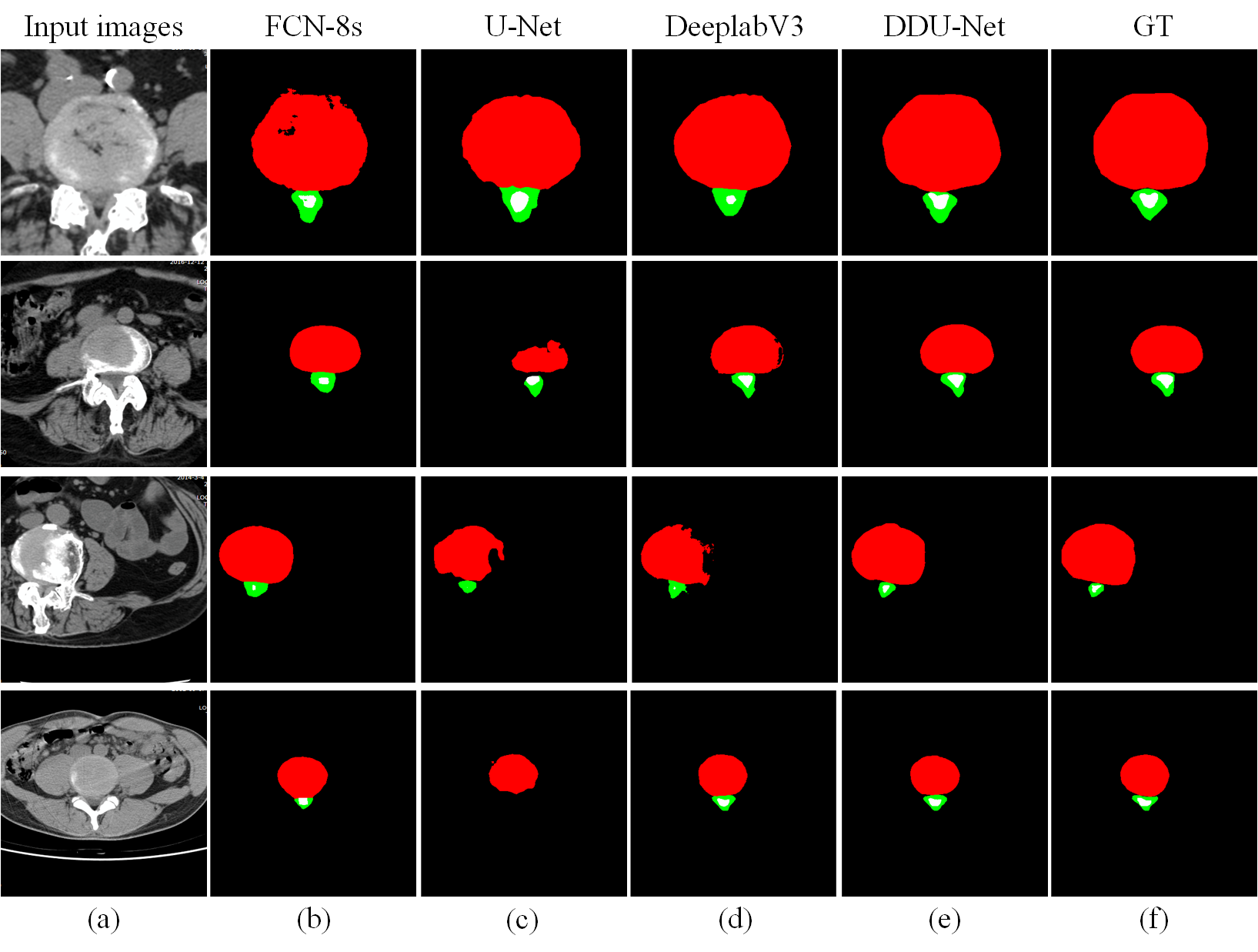}
{Visual comparison between DDU-Net and three state-of-the-art methods. The images in the rightmost column are the ground truth of each row, where the red regions indicate vertebral bodies, green regions indicate the spinal canal, white regions indicate the dural sac, and black regions indicate the background. From (b) to (e) are the results of FCN-8s \cite{long2015fully}, U-Net \cite{ronneberger2015u}, DeeplabV3 \cite{chen2017rethinking} and the proposed DDU-Net, respectively. Our results are the most close to the ground truth. \label{fig:8}}

\textbf{Quantitative Comparison}
The quantitative comparison of several methods on our dataset is shown in Table \ref{tab:5}. For fair comparison, the parameters of FCN-8s \cite{long2015fully}, U-Net \cite{ronneberger2015u} and DeeplabV3 \cite{chen2017rethinking} are finetuned on our dataset before comparing them. We can see that DDU-Net performs best in terms of both four evaluation metrics. Since over 95\% of the labels are the background class, the performances on the PA and fwIoU metrics are quite saturate, these two metrics are not appropriate to evaluate performance of one method. On the other hand, the performances of DDU-Net in terms of the mIoU and MPA metrics reach 0.8331 and 0.9099 respectively, surpassing the state-of-the-art methods by at least 3 points. Consequently, both qualitative and quantitative comparisons between our method and the state-of-the-art methods indicate that the proposed method can generate promising segmentations on practical lumbar spinal CT images.

\begin{table}[]
\caption{Quantitative comparison of different methods on our dataset. The best performances are bolded, and the second best performances are underlined.}
\centering
\small
\begin{tabular}{|l|l|l|l|l|}
\hline
Methods & mIoU & PA   & MPA   & fwIoU \\ \hline
FCN-8s \cite{long2015fully}             & 0.7705    & 0.9869    & 0.7588    & 0.9766 \\
U-Net \cite{ronneberger2015u}           & 0.7446    & 0.9804    & 0.7316    & 0.9646 \\
DeeplabV3 \cite{chen2017rethinking}     & \underline{0.8053}    & \underline{0.9896}    & \underline{0.8718}    & \underline{0.9802} \\
DDU-Net (proposed)                      & \textbf{0.8331}    &\textbf{0.9913}     &\textbf{0.9099} 	&\textbf{0.9835} \\ \hline
\end{tabular}
\label{tab:5}
\end{table}

\section{Conclusion}
\label{sec6}
Precisely identifying and recognizing the vertebral body, spinal canal and dural sac is a key step in diagnosing different types of LSS. In this paper, we first provide a new lumbar spinal CT image segmentation dataset with pixel-level labels and present a fully automatic method for segmentation of the vertebral body, spinal canal and dural sac from axial spine CT images based on a dual densely connected U-shaped network. Our method is practical, and requires no image preprocessing such as contrast enhancement, registration and denoising; the input is raw CT images, and the output is the desired segmentation maps; and the running speed is about 12 FPS (please see Table \ref{tab:3}). Our method is precise, and by comparing the segmentation results to those of existing state-of-the-art methods on our new dataset, the proposed method proved superior in terms of segmentation accuracy (Table \ref{tab:5}).

Given that we have automatically segmented the vertebral body, spinal canal and dural sac from CT images, there is still one more step before fully automatic LSS diagnosis of different types. In future work, we will apply the proposed DDU-Net as an approach for generating regions of interest and will investigate the complete automatic LSS diagnosis pipeline.


\EOD

\end{document}